\documentclass{article}
\usepackage{graphicx}

\usepackage{arxiv}

\usepackage[utf8]{inputenc}
\usepackage[T1]{fontenc}
\usepackage{hyperref}
\usepackage{url}
\usepackage{booktabs}
\usepackage{amsfonts}
\usepackage{nicefrac}
\usepackage{microtype}
\usepackage{xcolor}
\usepackage{multirow}
\usepackage{tabularx}
\usepackage{subcaption}
\usepackage{authblk}
\usepackage{multicol}
\usepackage{caption}

\title{Human-Centered AI in Multidisciplinary Medical Discussions: Evaluating the Feasibility of a Chat-Based Approach to Case Assessment.}

\author[1]{\textbf{Shinnosuke Sawano}\thanks{\texttt{sawanos-int@h.u-tokyo.ac.jp}}}
\author[1]{\textbf{Satoshi Kodera}\thanks{\texttt{KODERAS-INT@h.u-tokyo.ac.jp}}}

\affil[1]{Department of Cardiovascular Medicine, The University of Tokyo Hospital}

\begin{document}
\maketitle
\begin{abstract}
In this study, we investigate the feasibility of using a human-centered artificial intelligence (AI) chat platform where medical specialists collaboratively assess complex cases. As the target population for this platform, we focus on patients with cardiovascular diseases who are in a state of multimorbidity, that is, suffering from multiple chronic conditions. We evaluate simulated cases with multiple diseases using a chat application by collaborating with physicians to assess feasibility, efficiency gains through AI utilization, and the quantification of discussion content. We constructed simulated cases based on past case reports, medical errors reports and complex cases of cardiovascular diseases experienced by the physicians. The analysis of discussions across five simulated cases demonstrated a significant reduction in the time required for summarization using AI, with an average reduction of 79.98\%. Additionally, we examined hallucination rates in AI-generated summaries used in multidisciplinary medical discussions. The overall hallucination rate ranged from 1.01\% to 5.73\%, with an average of 3.62\%, whereas the harmful hallucination rate varied from 0.00\% to 2.09\%, with an average of 0.49\%. Furthermore, morphological analysis demonstrated that multidisciplinary assessments enabled a more complex and detailed representation of medical knowledge compared with single physician assessments. We examined structural differences between multidisciplinary and single physician assessments using centrality metrics derived from the knowledge graph. In this study, we demonstrated that AI-assisted summarization significantly reduced the time required for medical discussions while maintaining structured knowledge representation. These findings can support the feasibility of AI-assisted chat-based discussions as a human-centered approach to multidisciplinary medical decision-making.

\end{abstract}

\section{Introduction}\label{introduction}
Human-centered medical artificial intelligence (AI) research that explores AI integration in medical and envisions new clinical workflows is critically needed to alleviate the burden on medical professionals and enable seamless coexistence between humans and AI in medical practice (1-3). In recent years, large language models (LLMs) such as Chat Generative Pretrained Transformer (ChatGPT) have emerged as promising solutions and demonstrated their potential to enhance access to medical information for both medical professionals and patients by facilitating efficient knowledge retrieval, patient education, and clinical decision support (4-6). In studies evaluating the performance of ChatGPT on the United States Medical Licensing Examination (USMLE), researchers demonstrated that ChatGPT performed as well as medical students (7-9). However, the implementation of AI in medical practice has not progressed significantly. This is because medical environments are optimized individually, which makes AI integration potentially disruptive to existing workflows. Additionally, challenges persist, such as hallucination and high output variability. As a result, there are significant challenges in using generative AI such as ChatGPT independently for medical assessments (10). 

To address these challenges, in this study, we investigate the feasibility of using a human-centered chat platform where medical specialists assess complex cases collaboratively by leveraging AI to enhance efficiency and facilitate structured discussions. As the target population for this platform, we focus on patients with cardiovascular diseases who are in a state of multimorbidity, that is, suffering from multiple chronic conditions (11, 12). In previous studies, researchers reported that 64.9\% of individuals aged 65–84 and 81.5\% of those aged 85 and older experience multimorbidity (13). This poses significant challenges in medical care and service provision because it affects not only individual diseases but also the interactions between them, which effects patients' health, quality of life, and medical resource utilization (14). Patients with multimorbidity often receive treatment from multiple medical specialists, which leads to a lack of care coordination and increasing concerns about adverse events caused by overlapping or interacting pharmacotherapies (15). Given this background, there is a growing need to establish a new healthcare system that provides comprehensive assessments for patients with multimorbidity.

In this study, we evaluate simulated cases with multiple diseases using a chat application with physicians and assess its feasibility, the efficiency gains achieved through AI utilization, and the quantification of discussion content.

\section{Method}
\subsection{Study design}\label{Study design}
We reviewed chat discussions among physicians in simulated cases. We constructed the simulated cases based on past case reports, incident reports of medical errors, and complex cases of cardiovascular diseases experienced by physicians who did not participate in the discussions. Two physicians reviewed and double-checked these cases to ensure medical validity. We selected the physicians included in the discussions based on their expertise and clinical experience in providing comprehensive evaluations for patients with multimorbidity, including cardiovascular diseases. The eligibility criteria included being board-certified specialists in cardiology, nephrology, diabetology, or general medicine, or primary care physicians, and a minimum of five years of clinical experience post-medical licensure. Additionally, the physicians were required to fully understand the objectives of this study and the significance of the AI-assisted chat platform, and provide informed consent for participation.

\subsection{Workflow of Generative AI-Assisted Case Discussion}
Discussions were conducted on the generative AI chat platform based on the medical information of the simulated cases. We used ChatGPT-4o as the generative AI model (16). We assigned specialists, cardiologists, nephrologists, diabetologists, and primary care physicians to each case. We informed participating physicians in advance that the cases were fictional. The assigned specialists recorded the physicians medical evaluations on the platform based on the medical information provided for the simulated cases. Once all participating physicians completed their evaluations, we summarized the content using generative AI with custom prompts. Physicians at our institution reviewed and revised the validity of the AI-generated summaries as necessary while simultaneously verifying for hallucinations.

\subsection{Evaluation of the Impact of Generative AI}
To evaluate efficiency, we measured the time required to generate the final reports, focusing on the impact of AI-generated summaries on the overall report creation process. To maintain consistency in processing, we used a standardized format for the reports. Specifically, we compared the time required from the AI-generated summary to the finalized report with the time required to create the report without AI assistance. We calculated the time based on the final report’s word count, assuming a writing speed of 60 characters per minute.

 \subsection{Review of Hallucinations in AI-Generated Text by Physicians}
To evaluate hallucinations in AI-generated summaries, we conducted morphological analysis on pre-finalized text reviewed by physicians (17, 18). In other words, the raw output generated by AI was directly reviewed by physicians. For morphological analysis, we used MeCab with the Neologd dictionary (https://github.com/neologd/mecab-ipadic-neologd/blob/master/README.ja.md), which is widely used for Japanese text processing (17). The AI-generated summaries were analyzed morphologically, and the physicians identified and counted all instances of expressions that failed to convey the intended meaning accurately, clearly erroneous descriptions, and other hallucinations through manual verification.

Furthermore, among all hallucinations, we classified those considered medically unacceptable as "harmful hallucinations." We further categorized harmful hallucinations based on whether they occurred at the word level or sentence level. For quantification, we calculated the error rate by determining the proportion of all hallucinations or harmful hallucinations relative to the total number of morphological tokens extracted per case.

 \subsection{Knowledge Graph-Based Comparison of Multidisciplinary Teams and Single Physician Assessments}
For quantitative evaluation in English, we systematically organized the medical knowledge generated from the discussions into a structured knowledge graph, and integrated both AI-generated summaries based on multidisciplinary discussions reviewed by physicians and summaries independently created by a single physician. This approach facilitated the integration of both objective and subjective aspects of clinical reasoning into a unified, analyzable framework. To ensure consistency and accuracy in terminology, we primarily used SNOMED Clinical Terms® (SNOMED CT®), which is a globally recognized standard for medical vocabulary (www.snomed.org). In the knowledge graph, two nodes (e.g., diseases, findings, or other clinical concepts) were connected by an edge, which represented the relationship between them (e.g., "associated finding" or "has interpretation"). For concepts or relationships that could not be expressed adequately using SNOMED CT, we used custom terminology developed specifically for this study. 
 
In this study, we did not aim to develop a universally applicable medical knowledge graph. Instead, we tailored the knowledge graph to a closed environment for each case. It captured the specific language and reasoning used by physicians during the discussions to describe the pathophysiology and management of individual cases. By focusing on patient-specific contexts, the knowledge graph ensured an accurate representation of nuanced clinical decision-making. For analysis, we compared the structural complexity of knowledge graphs generated from evaluations conducted by a single physician versus those conducted through multidisciplinary discussions. Specifically, we assessed the number of branches, number of leaf nodes, maximum leaf node depth, and average leaf node depth to quantify the hierarchical depth and breadth of medical discussions. The number of branches represents the total number of decision pathways emerging from the initial clinical discussion, which reflects the diversity of diagnostic or treatment considerations. The number of leaf nodes indicates the total number of terminal points within the knowledge graph, which represents conclusions or final decision points. The maximum leaf node depth measures the deepest hierarchical level in the network, which signifies the most extended logical reasoning chain. The average leaf node depth provides an overall assessment of how deep the discussions generally extended across cases.

Additionally, we analyzed centrality metrics, including degree centrality, betweenness centrality, closeness centrality, and eigenvector centrality, to identify key medical knowledge nodes within the discussions. Degree centrality assesses the extent of the direct connections a node had, which indicates its prominence in the network. Betweenness centrality measured the extent to which a node acted as a bridge in the shortest paths between other nodes, thereby highlighting critical intermediary nodes. Closeness centrality quantified how efficiently information could spread within the network, whereas eigenvector centrality identified nodes that were highly influential because of their connections to other highly connected nodes.

 \subsection{Statistical analysis}
Continuous data are expressed as the mean ± standard deviation. As normal, we compared distributed continuous variables using the Student’s t-test and non-normally distributed continuous variables using the Mann–Whitney U test. We performed statistical analysis using JMP Pro 18 (SAS Institute, Tokyo, Japan) and defined statistical significance as P-value < 0.05.

\section{Results}
\begin{figure}[ht]
  \centering
  \includegraphics[width = 1.0\linewidth]{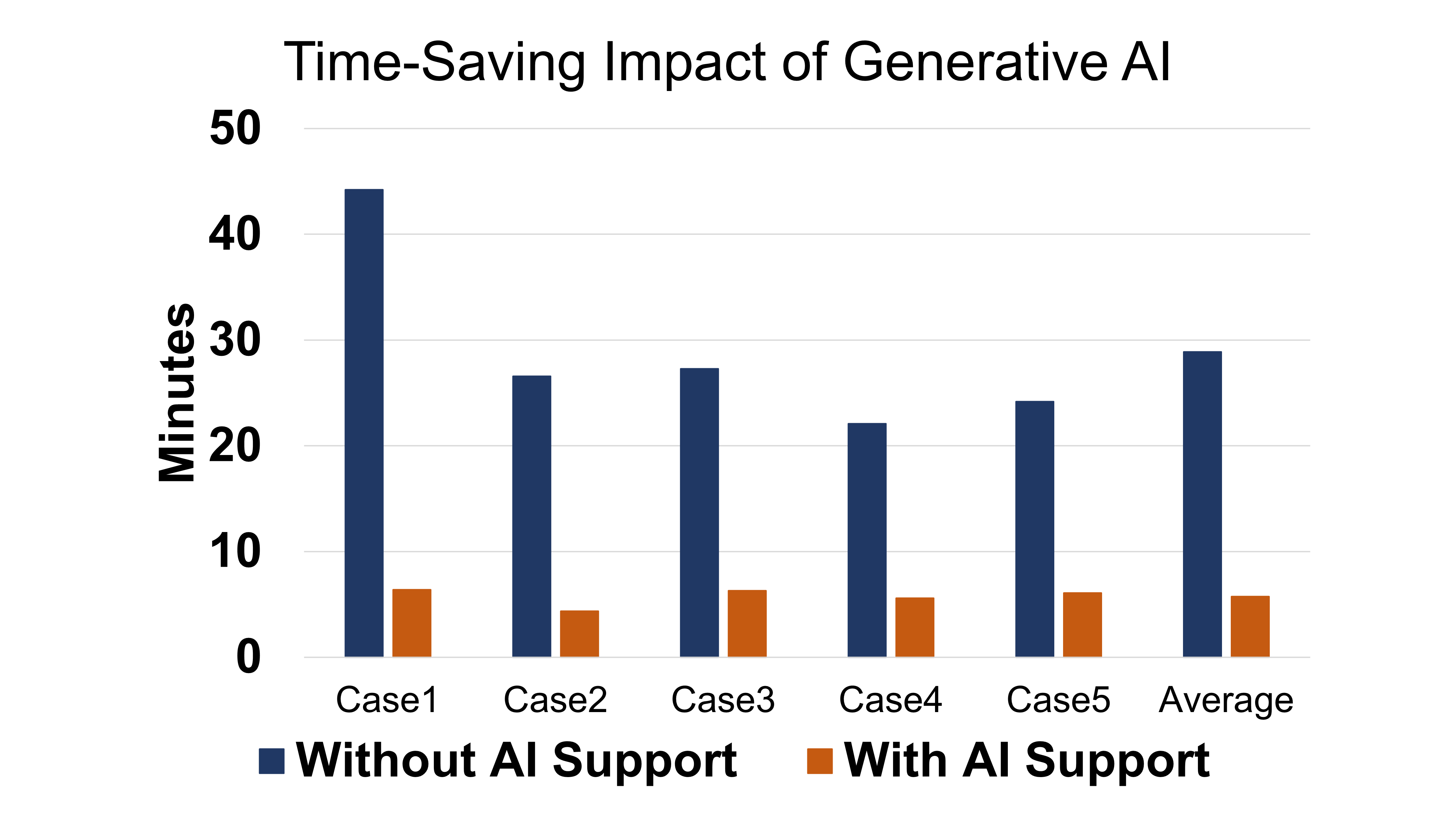}
  \caption{Time-Saving Impact of Generative AI. Illustration of the time-saving impact of generative AI in case discussions using a comparison of the time required for case assessments with and without AI support. The results demonstrated that AI-assisted discussions led to a 79.98\% reduction in the time required for case assessment.}
  \label{Figure 1}
\end{figure}

The analysis of discussions across five simulated cases demonstrated a significant reduction in the time required for summarization using AI. The total discussion time spanned an average of 28 minutes and 58 seconds (28.98 minutes) across the cases without AI support, whereas the AI-assisted summarization process took only 5 minutes and 48 seconds (5.8 minutes) on average. This corresponds to a 79.98\% reduction in time (Figure~\ref{Figure 1}). We based the calculation on the average word count of 1738.6 characters per case in Japanese, which contributed to determining these results. 

\begin{figure}[ht]
  \centering
  \includegraphics[width = 1.0\linewidth]{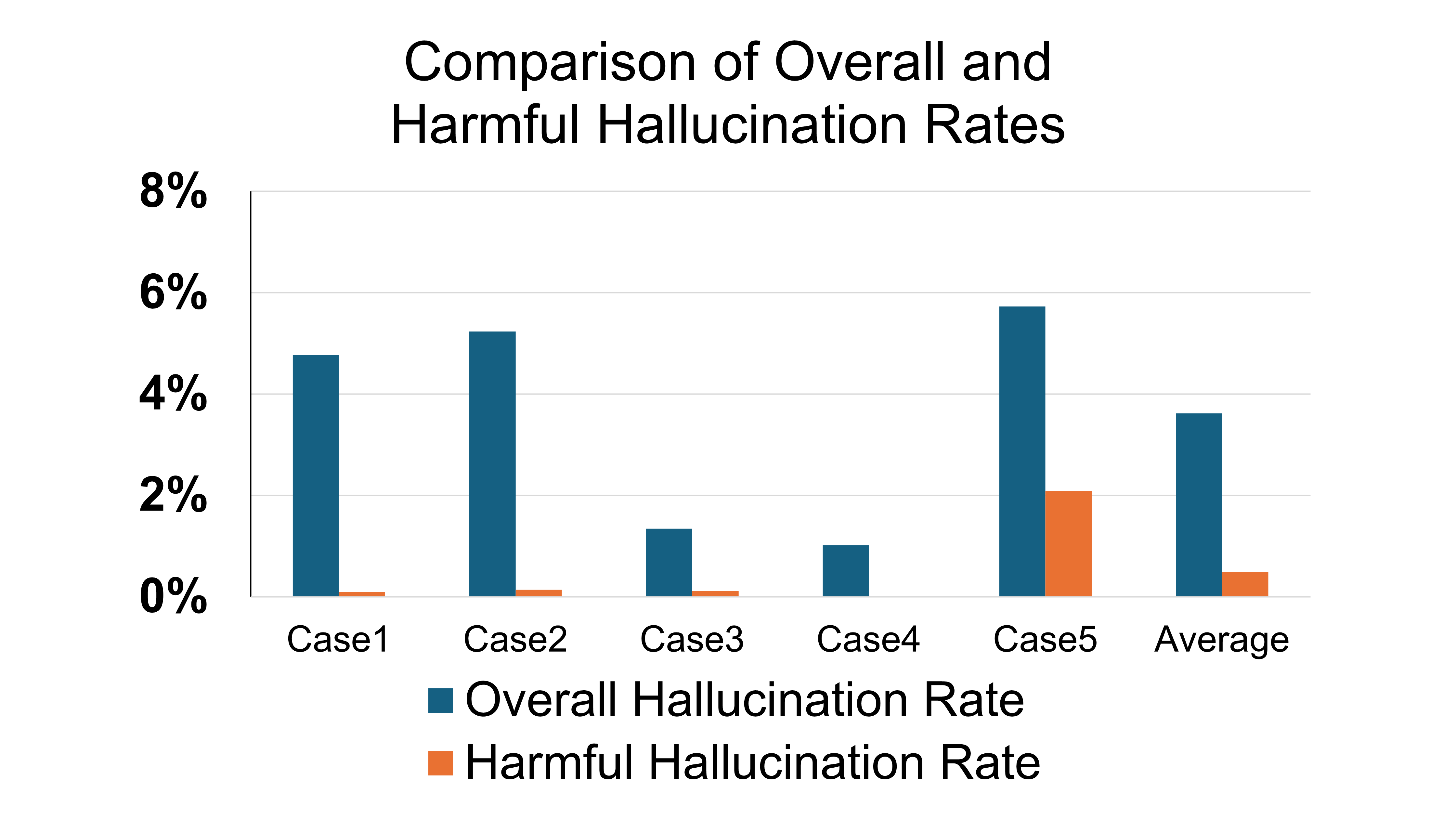}
  \caption{Comparison of Overall and Harmful Hallucination Rates. Illustration of the hallucination rates observed in AI-generated summaries used in multidisciplinary medical discussions. The overall hallucination rate ranged from 1.01\% to 5.73\%, with an average of 3.62\%. The harmful hallucination varied from 0.00\% to 2.09\%, with an average of 0.49\%.}
  \label{Figure 2}
\end{figure}

Tables~\ref{table1} provides a detailed breakdown of the hallucinations observed in each case. The total number of morphological tokens per case ranged from 690 to 1070. The number of inaccurate descriptions detected varied between 7 and 51, whereas the number of harmful descriptions ranged from 0 to 15. Notably, we categorized harmful hallucinations based on their occurrence at either the word level or sentence level. In three cases, harmful hallucinations appeared as single words, whereas in Case 5, they predominantly appeared at the sentence level (1 sentence, 15 tokens). Figure~\ref{Figure 2} shows the overall hallucination rate and harmful hallucination rate for each simulated patient case. The overall hallucination rate ranged from 1.01\% to 5.73\%, with an average of 3.62\%. The harmful hallucination rate varied from 0.00\% to 2.09\%, with an average of 0.49\%.

\begin{table}[!ht]
  \centering
  \captionsetup{skip=10pt}
  \begin{tabularx}{\textwidth}{X X X X >{\raggedright\arraybackslash}X} 
    \toprule
    Variables & Total Morphological Tokens & Inaccurate Hallucinated Tokens
      & Harmful Hallucinated Tokens & Type of Harmful Hallucination \\
    \midrule    
    Case1 & 1070 & 51 & 1  & 1 word \\
    Case2 & 745  & 39 & 1  & 1 word \\
    Case3 & 893  & 12 & 1  & 1 word \\
    Case4 & 690  & 7  & 0  & - \\
    Case5 & 716  & 41 & 15 & 1 sentence \\
    \bottomrule
  \end{tabularx}
    \caption{Breakdown of the Hallucinations Observed in Each Case. The total number of morphological tokens represents the total linguistic elements obtained by analyzing each AI-generated summary using MeCab. Inaccurate hallucinated tokens refer to instances where the AI-generated text failed to convey the intended meaning correctly. Harmful hallucinated tokens are a subset of inaccurate hallucinated tokens that were deemed medically unacceptable based on expert review. The type of harmful hallucination is categorized based on whether the error occurred at the word level or sentence level. Cases without harmful hallucinations are marked with a hyphen (-).} 
    \label{table1}
\end{table}

The number of branches was significantly higher in multidisciplinary assessments compared with single physician assessments (48.0 ± 3.2 vs. 27.4 ± 3.2, p = 0.002), which suggests a more complex and detailed representation of medical knowledge. The maximum leaf node depth was also significantly greater in multidisciplinary assessments (6.0 ± 0.4 vs. 3.2 ± 0.4, p = 0.001), which indicates deeper hierarchical structures. The average leaf node depth followed a similar trend, with a significantly greater depth in multidisciplinary assessments (3.2 ± 1.2 vs. 2.3 ± 0.6, p < 0.001). By contrast, the number of leaf nodes did not differ significantly between the two groups (16.4 ± 1.1 vs. 14.2 ± 1.1, p = 0.193). These results were based on five simulated patient cases to ensure consistency in case complexity across both assessment types (Tables~\ref{table2}).

\begin{table}[!ht]
  \centering
  \captionsetup{skip=10pt}
  \begin{tabularx}{\textwidth}{X X X X X} 
    \toprule
    Variables & Multidisciplinary & Single Physician & P value \\
    \midrule    
    Number of Branches & 48.0 ± 3.2 & 27.4 ± 3.2 & 0.002\\
    Number of Leaf Nodes & 16.4 ± 1.1  & 14.2 ± 1.1 & 0.193 \\
    Max Leaf Node Depth & 6.0 ± 0.4  & 3.2 ± 0.4 & 0.001 \\
    Average Leaf Node Depth & 3.2 ± 1.2  & 2.3 ± 0.6  & < 0.001 \\
    \bottomrule
  \end{tabularx}
    \caption{Comparison of the Branch and Leaf Node Depth between multidisciplinary teams and Single physician assessments. Comparative analysis of the hierarchical complexity in knowledge graphs generated from five simulated patient cases assessed by multidisciplinary teams and single physicians. The values are expressed as mean ± standard deviation (SD). P-values were calculated using an independent t-test or Mann-Whitney U test, depending on the data distribution.} 
    \label{table2}
\end{table}

Tables~\ref{table3} shows the results of centrality metrics between multidisciplinary and single physician assessments. The degree centrality was significantly higher in single physician assessments compared with multidisciplinary assessments (0.084 ± 0.062 vs. 0.058 ± 0.043, p < 0.001), which indicates a more direct and centralized knowledge structure. By contrast, betweenness centrality was significantly greater in multidisciplinary assessments (0.012 ± 0.030 vs. 0.002 ± 0.002, p < 0.001), which suggests that multidisciplinary teams rely more on intermediary nodes to connect different medical concepts. To facilitate a deeper understanding of the structural differences in knowledge representation, we provide a comparative visualization of degree centrality and betweenness centrality in Supplemental Figure~\ref{SF1}and\ref{SF2}. These figures illustrate how centrality measures influence the structure of multidisciplinary discussions and single physician assessments. Closeness centrality showed no significant difference between the two groups (0.056 ± 0.026 vs. 0.055 ± 0.026, p = 0.541), which indicates similar overall accessibility of information within the networks. Similarly, eigenvector centrality did not differ significantly between multidisciplinary and single physician assessments (0.067 ± 0.141 vs. 0.068 ± 0.179, p = 0.947), which suggests that the influence of highly connected nodes remained comparable across both assessment types.

\begin{table}[!ht]
  \centering
  \captionsetup{skip=10pt}
  \begin{tabularx}{\textwidth}{X X X X X} 
    \toprule
    Variables & Multidisciplinary & Single Physician & P value \\
    \midrule    
    Degree Centrality & 0.058 ± 0.043 & 0.084 ± 0.062 & < 0.001\\
    Betweenness Centrality & 0.012 ± 0.030  & 0.002 ± 0.002 & < 0.001 \\
    Closeness Centrality & 0.056 ± 0.026  & 0.055 ± 0.026 & 0.541 \\
    Eigenvector Centrality & 0.067 ± 0.141  & 0.068 ± 0.179  & 0.947 \\
    \bottomrule
  \end{tabularx}
    \caption{Comparison of Centrality Measures between Multidisciplinary and Single Physician Assessments. Comparative analysis of centrality measures in knowledge graphs generated from five simulated patient cases assessed by multidisciplinary teams and single physicians. The values are expressed as mean ± standard deviation (SD). P-values were calculated using the Mann–Whitney U test to account for a potential non-normal distribution. Degree centrality represents the number of direct connections each node has in the network. Betweenness centrality quantifies the extent to which a node acts as a bridge between other nodes. Closeness centrality measures the average distance from a node to all other nodes in the network. Eigenvector centrality reflects a node’s influence based on its connections to other highly connected nodes. P-value < 0.05 was considered statistically significant.} 
    \label{table3}
\end{table}

Data Availability: The data are not publicly available because this study includes valuable simulated cases created by specialists. Given the recent trend of LLMs using vast amounts of data for training, the importance of maintaining closed-case data is increasing. Data are available from the Cardiovascular Medicine Department at the University of Tokyo Hospital (contact via kayo.cho@gmail.com, who is the person in charge of data management at the Department of Cardiovascular Medicine, University of Tokyo Hospital) for researchers who meet the criteria for access to confidential data.

\section{Discussion}
The results of this study highlight the potential of AI-assisted summarization in reducing the time required for synthesizing medical discussions while maintaining a structured and efficient knowledge representation. The implementation of AI reduced the summarization time by approximately 80\%, which significantly enhanced workflow efficiency. However, the analysis of hallucinations suggests that, although hallucinations were present, effective role-sharing between AI and human experts enable the delivery of more efficient medical assessments to patients. Particularly in multidisciplinary assessments, the complexity of branching structures and node depth was higher, which allowed for a more comprehensive representation of knowledge. Centrality metrics further demonstrated distinct patterns in knowledge distribution between single physician and multidisciplinary assessments, which emphasized the role of collaborative discussions in medical decision-making. Overall, in this study, we demonstrated the potential of human-AI collaboration to provide personalized medical information to patients.

We should discuss the importance of AI-assisted summarization in reducing the time required for chat-based medical discussions. First, we demonstrated that it is feasible to gather specialists and conduct discussions on a chat platform using simulated patient information. This indicates the potential for physicians to provide medical assessments to patients without being constrained by temporal or spatial limitations. However, because of the nature of chat-based discussions, specialist contributions tend to be scattered, thereby requiring significant time for summarization. In fact, the discussions for the five simulated cases in this study took an average of 28.98 minutes. By contrast, AI-assisted summarization reduced this processing time by 79.98\%. This suggests that AI can enhance the feasibility of using chat-based platforms for the provision of medical expertise by specialists. As a result, the implementation of AI-assisted summarization could contribute to maximizing the utilization of medical knowledge and resources, ultimately leading to more efficient healthcare delivery.

We should also analyze the hallucinations observed in each case. All five cases exhibited some level of hallucination, and in four cases, physicians identified some of these as harmful hallucinations. Although a strict comparison of hallucination rates is challenging because of the lack of studies under identical conditions, our findings suggest that the observed hallucination rates align with or are lower than those reported in previous research on AI-generated medical summaries (19, 20). In a manual review, physicians observed that patient evaluations based on broader discussions, such as best supportive care or considerations of the patient's values, tended to have fewer hallucinations. Notably, Case 4, which had no harmful hallucinations, followed this discussion style. By contrast, when discussions were focused on specific diseases or detailed intervention methods, as seen in the other four cases, hallucinations occurred at both the word and sentence levels. Although improvements to the model’s specialized medical knowledge are needed to reduce hallucinations (21, 22), it provides a lower bound on hallucination rates, which suggests that hallucinations are unavoidable to some extent, even in ideal conditions with perfect training data (23). Therefore, in this study, we emphasized the importance of collaboration between AI and physicians. In this division of labor, AI can efficiently summarize information to reduce the burden on physicians while physicians ensure the accuracy of the information before they deliver it to patients. This approach has the potential to provide patients with higher-quality medical information in a more efficient manner.

The results of this study suggest that multidisciplinary assessments enable a more complex and detailed representation of medical knowledge compared with single physician assessments. Specifically, the number of branches was significantly higher in multidisciplinary assessments. Additionally, both the maximum and average leaf node depths were significantly greater, which indicates that multidisciplinary assessments possess a deeper hierarchical structure and facilitate more detailed and in-depth discussions. By contrast, no significant difference was observed in the total number of leaf nodes between the two groups. This result implies that, although multidisciplinary assessments increase branching and depth, the final number of conclusions remains relatively unchanged; that is, even in single physician assessments, the numbers of final diagnoses and decisions do not differ significantly. Instead, multidisciplinary assessments are likely to contribute to a more enriched and comprehensive discussion, thereby enhancing the overall quality of the evaluation rather than simply increasing the number of conclusions.

We compared centrality metrics between multidisciplinary and single physician assessments, which identified the structural characteristics of the knowledge graph. In single physician assessments, degree centrality was significantly higher, which suggests a more direct and centralized knowledge structure. This implies that a single physician makes rapid decisions, which leads to a simplified flow of information.

By contrast, betweenness centrality was significantly higher in multidisciplinary assessments. This indicates that physicians from different specialties play a complementary role in connecting information. By integrating insights from various perspectives and linking multiple concepts, multidisciplinary discussions may lead to more comprehensive evaluations. Furthermore, we observed no significant differences in closeness centrality and eigenvector centrality between multidisciplinary and single physician assessments. This result suggests that, despite the increased branching in multidisciplinary discussions, the overall efficiency of information transmission remained similar across both groups. Additionally, the influence of key informational nodes appeared to be evenly distributed, which indicates that no single expert dominated the discussion but rather, a balanced exchange of knowledge was maintained in both settings

This study has the following limitations. First, we conducted this study using simulated cases and did not perform validation with real patients. As a result, the feasibility of chat-based discussions cannot be directly applied to real clinical settings. Additionally, we limited the number of cases analyzed to five and also restricted the number of discussion sessions. This inherently introduced bias in the morphological analysis and knowledge graph structure evaluations because of the small sample size. We used SNOMED CT® to standardize medical terminology in the knowledge graph, but several limitations exist. The conversion of Japanese medical expressions into English potentially introduced inaccuracies. Then, not all medical expressions could be fully integrated into the knowledge graph, which may have resulted in researcher-induced biases in the graph construction. Furthermore, we constructed the knowledge graph within a closed environment for each case, which makes its generalizability difficult. Moreover, we used only ChatGPT-4o as the generative AI model and did not conduct comparisons with other models. This limits the generalizability of the findings to different LLMs and may not fully capture the variations in AI performance across models. Finally, we did not evaluate the effect of multidisciplinary assessments on patient outcomes. In future studies, we should investigate how multidisciplinary discussions influence clinical decision-making, treatment effectiveness, and patient prognosis to assess the real-world benefits of AI-assisted summarization.

\section{Conclusions}
In this study, we demonstrated that AI-assisted summarization significantly reduced the time required for medical discussions while maintaining structured knowledge representation. In future research, we should focus on real-world validation and optimizing human-centered AI collaboration to enhance clinical decision-making and patient outcomes.

\section{Acknowledgements}
We would like to express our sincere gratitude to Shusaku Egami for his invaluable contributions to the application of knowledge graphs. We also extend our appreciation to Risa Kishikawa, Masataka Sato, Hitoshi Naito, and Takaaki Koide for their expertise and support in the clinical evaluation of patients. Furthermore, we thank Thomas Kenta Cameron for his dedicated efforts in application development. Their insights and assistance have been instrumental in the completion of this work.

We also thank Edanz (https://jp.edanz.com/ac) for editing a draft of this manuscript.

\section{Ethics approval}
This study was approved by the Institutional Review Board of The University of Tokyo (reference number 2024502NI).

\section*{References}

\medskip

{
\small

1.	Thieme A, Rajamohan A, Cooper B, Groombridge H, Simister R, Wong B, et al. Challenges for responsible AI design and workflow integration in healthcare: a case study of automatic feeding tube qualification in radiology. ACM Transactions on Computer-Human Interaction. 2024.

2.	Beede E, Baylor E, Hersch F, Iurchenko A, Wilcox L, Ruamviboonsuk P, et al., editors. A human-centered evaluation of a deep learning system deployed in clinics for the detection of diabetic retinopathy. Proceedings of the 2020 CHI conference on human factors in computing systems; 2020.

3.	Burgess ER, Jankovic I, Austin M, Cai N, Kapuścińska A, Currie S, et al. Healthcare AI Treatment Decision Support: Design Principles to Enhance Clinician Adoption and Trust.  Proceedings of the 2023 CHI Conference on Human Factors in Computing Systems; Hamburg, Germany: Association for Computing Machinery; 2023. p. Article 15.

4.	Wei Q, Yao Z, Cui Y, Wei B, Jin Z, Xu X. Evaluation of ChatGPT-generated medical responses: A systematic review and meta-analysis. J Biomed Inform. 2024;151:104620.

5.	Lyu Q, Tan J, Zapadka ME, Ponnatapura J, Niu C, Myers KJ, et al. Translating radiology reports into plain language using ChatGPT and GPT-4 with prompt learning: results, limitations, and potential. Vis Comput Ind Biomed Art. 2023;6(1):9.

6.	Fleming SL, Lozano A, Haberkorn WJ, Jindal JA, Reis E, Thapa R, et al., editors. Medalign: A clinician-generated dataset for instruction following with electronic medical records. Proceedings of the AAAI Conference on Artificial Intelligence; 2024.

7.	Kung TH, Cheatham M, Medenilla A, Sillos C, De Leon L, Elepaño C, et al. Performance of ChatGPT on USMLE: Potential for AI-assisted medical education using large language models. PLOS Digit Health. 2023;2(2):e0000198.

8.	Nori H, King N, McKinney SM, Carignan D, Horvitz E. Capabilities of gpt-4 on medical challenge problems. arXiv preprint arXiv:230313375. 2023.

9.	Brin D, Sorin V, Vaid A, Soroush A, Glicksberg BS, Charney AW, et al. Comparing ChatGPT and GPT-4 performance in USMLE soft skill assessments. Scientific Reports. 2023;13(1):16492.

10.	Freyer O, Wiest IC, Kather JN, Gilbert S. A future role for health applications of large language models depends on regulators enforcing safety standards. Lancet Digit Health. 2024;6(9):e662-e72.

11.	Boyd CM, Darer J, Boult C, Fried LP, Boult L, Wu AW. Clinical practice guidelines and quality of care for older patients with multiple comorbid diseases: implications for pay for performance. Jama. 2005;294(6):716-24.

12.	Marengoni A, Angleman S, Melis R, Mangialasche F, Karp A, Garmen A, et al. Aging with multimorbidity: a systematic review of the literature. Ageing research reviews. 2011;10(4):430-9.
13.	Barnett K, Mercer SW, Norbury M, Watt G, Wyke S, Guthrie B. Epidemiology of multimorbidity and implications for health care, research, and medical education: a cross-sectional study. The Lancet. 2012;380(9836):37-43.

14.	Van den Akker M, Buntinx F, Metsemakers JF, Roos S, Knottnerus JA. Multimorbidity in general practice: prevalence, incidence, and determinants of co-occurring chronic and recurrent diseases. Journal of clinical epidemiology. 1998;51(5):367-75.

15.	Boyd CM, Darer J, Boult C, Fried LP, Boult L, Wu AW. Clinical practice guidelines and quality of care for older patients with multiple comorbid diseases: implications for pay for performance. Jama. 2005;294(6):716-24.

16.	Hurst A, Lerer A, Goucher AP, Perelman A, Ramesh A, Clark A, et al. Gpt-4o system card. arXiv preprint arXiv:241021276. 2024.

17.	Kudo T, Yamamoto K, Matsumoto Y, editors. Applying conditional random fields to Japanese morphological analysis. Proceedings of the 2004 conference on empirical methods in natural language processing; 2004.

18.	Ji Z, Lee N, Frieske R, Yu T, Su D, Xu Y, et al. Survey of hallucination in natural language generation. ACM Computing Surveys. 2023;55(12):1-38.

19.	Umapathi LK, Pal A, Sankarasubbu M, editors. Med-HALT: Medical Domain Hallucination Test for Large Language Models. Conference on Computational Natural Language Learning; 2023.

20.	Vishwanath PR, Tiwari S, Naik TG, Gupta S, Thai DN, Zhao W, et al., editors. Faithfulness Hallucination Detection in Healthcare AI. Artificial Intelligence and Data Science for Healthcare: Bridging Data-Centric AI and People-Centric Healthcare; 2024.

21.	Bassamzadeh N, Methani C. A Comparative Study of DSL Code Generation: Fine-Tuning vs. Optimized Retrieval Augmentation. arXiv preprint arXiv:240702742. 2024.

22.	Achiam J, Adler S, Agarwal S, Ahmad L, Akkaya I, Aleman FL, et al. Gpt-4 technical report. arXiv preprint arXiv:230308774. 2023.

23.	Kalai AT, Vempala SS, editors. Calibrated language models must hallucinate. Proceedings of the 56th Annual ACM Symposium on Theory of Computing; 2024.

\newpage
\appendix
\section*{Appendix}

\captionsetup[figure]{labelformat=simple, labelsep=colon, name={Supplemental Figure}}
\setcounter{figure}{0}
\section{Supplemental Figure 1}
\begin{figure}[ht]
  \centering
  \includegraphics[width = 1.0\linewidth]{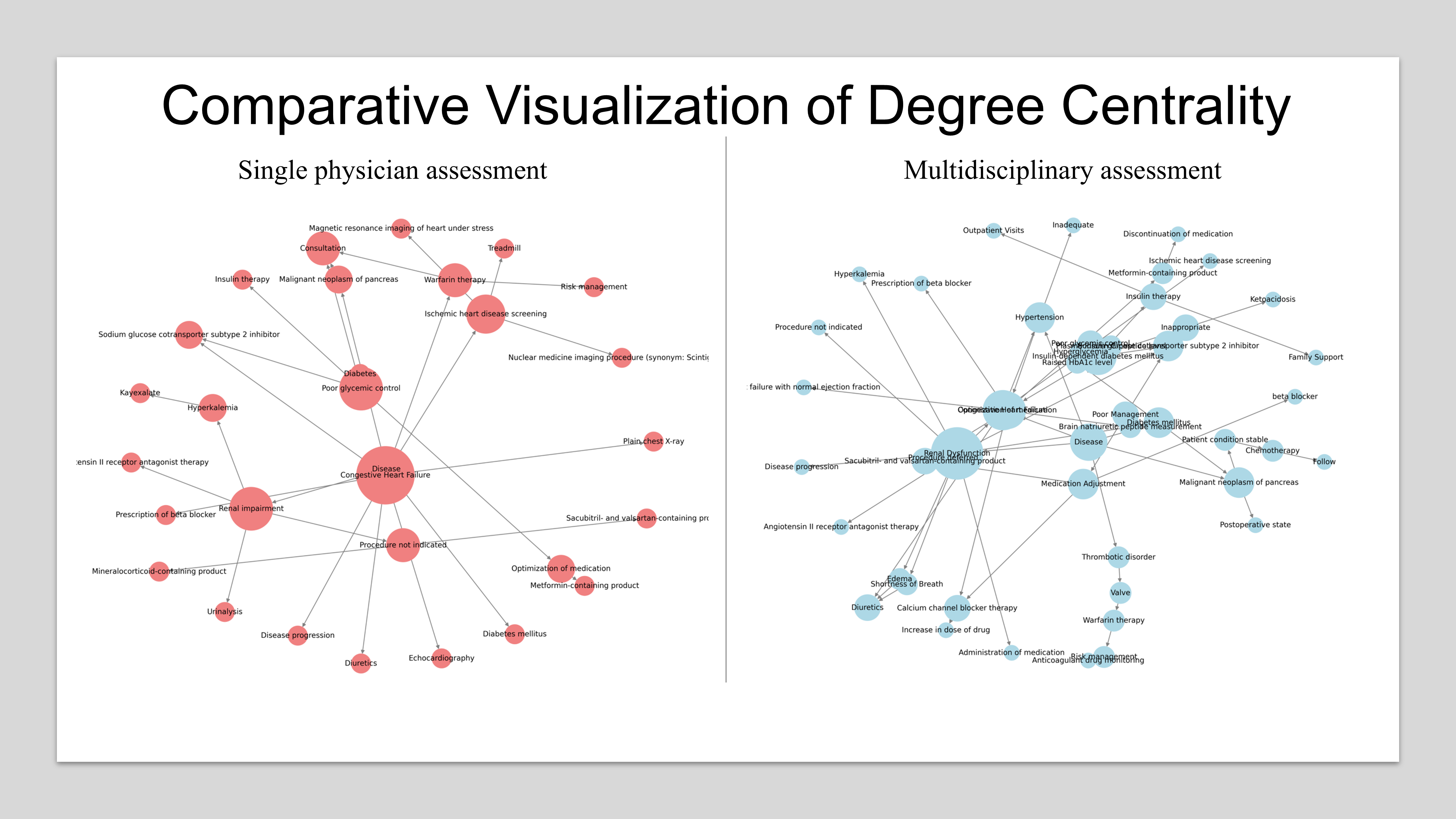}
  \caption{Visualization of Degree Centrality for Comparative Analysis.
Degree centrality visualization highlights nodes with a high number of direct connections, thereby representing the prominence of specific medical concepts within discussions. Degree centrality was significantly higher in single physician assessments (0.084 ± 0.062) compared with multidisciplinary discussions (0.058 ± 0.043), with a p-value of < 0.001. To facilitate visual comparison, the degree centrality values for both groups were scaled by the same fixed factor before node visualization, ensuring that differences between the two groups are more clearly distinguishable.
}
  \label{SF1}
\end{figure}

\newpage
\section{Supplemental Figure 2}
\begin{figure}[ht]
  \centering
  \includegraphics[width = 1.0\linewidth]{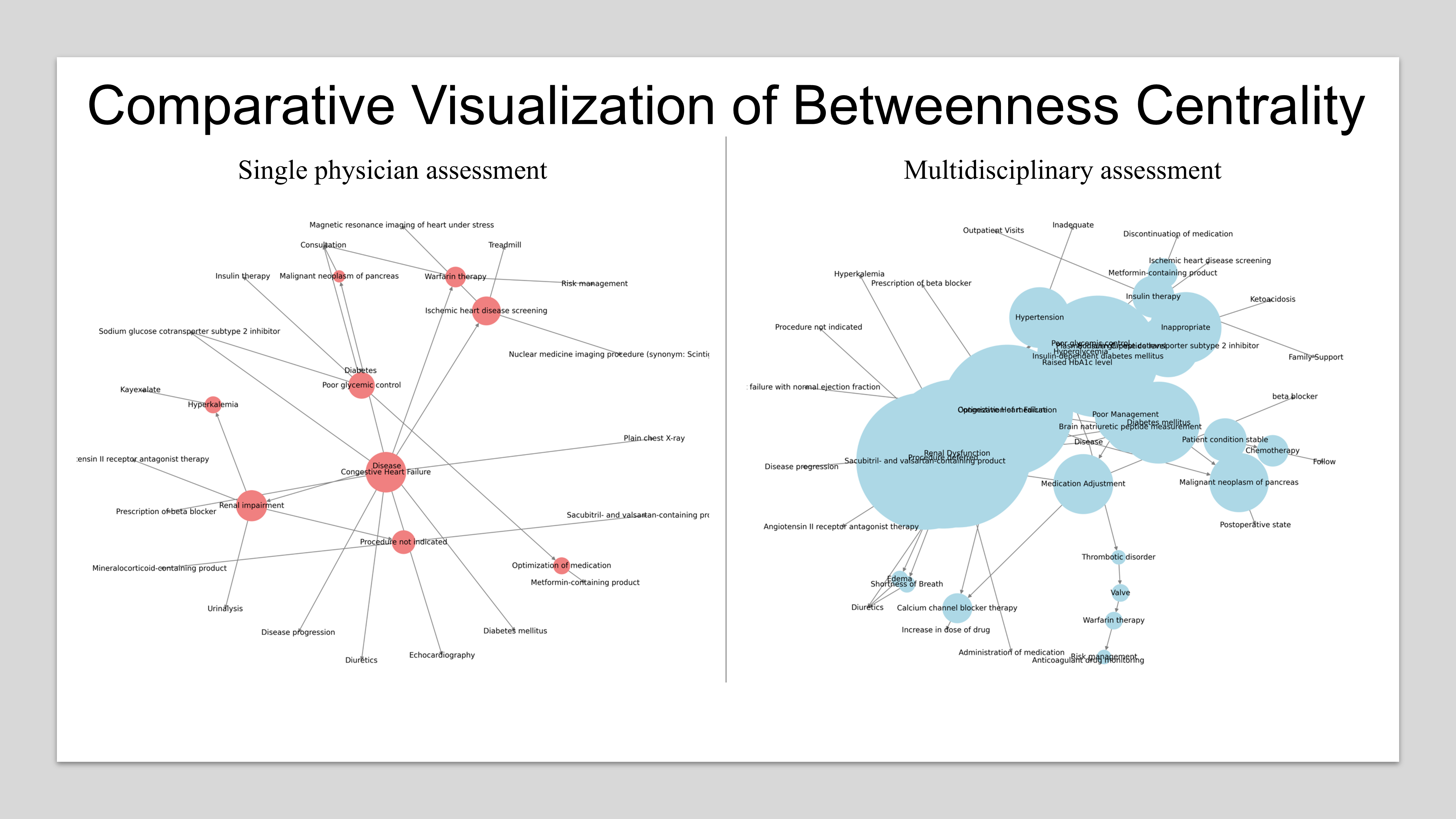}
  \caption{Visualization of Betweenness Centrality for Comparative Analysis.
Betweenness centrality visualization identifies key intermediary nodes that act as bridges in the knowledge network, thereby reflecting the flow of clinical reasoning and decision-making pathways. Betweenness centrality was significantly higher in multidisciplinary teams (0.012 ± 0.030) compared with single physician assessments (0.002 ± 0.002), with a p-value of < 0.001. To facilitate visual comparison, the betweenness centrality values for both groups were scaled by the same fixed factor before node visualization, ensuring that differences between the two groups are more clearly distinguishable.
}
  \label{SF2}
\end{figure}

\end{document}